\documentclass[prl,twocolumn,showpacs]{revtex4}

\usepackage{color,graphicx}
\usepackage{dcolumn}
\usepackage{bm}

\def\bfq {{\bf q}}

\def\bfK{{\bf K}}

\def\bfb{{\bf b}}
\def\bfp{{\bf p}}

\def\be{\begin{equation}}
 \def \ee{\end{equation}}
\def\bea{\begin{eqnarray}}
  \def\eea{\end{eqnarray}}

\def\bfb{{\bf b}}
\begin{document}

\title{
\vspace*{-1cm}
\begin{flushright}
{{\small \sl Version 20.1(\today)}}
\end{flushright}
\vspace*{-0.3cm}
Quantum Opacity, the RHIC HBT Puzzle, and the Chiral Phase Transition}

\author{John G. Cramer, Gerald A. Miller, Jackson M. S. Wu}
\author{Jin-Hee Yoon}\altaffiliation{Inha University, Republic of Korea.}
\affiliation{Department of Physics, University of Washington\\
  Seattle, WA 98195-1560}

\date{\today}

\begin{abstract}
We present a relativistic quantum mechanical treatment of opacity and refractive effects 
that allows reproduction of observables measured in two-pion 
(HBT) interferometry and pion spectra at RHIC. The inferred emission duration 
is substantial.
 The 
results are
consistent with the emission of pions from a system that has a restored chiral symmetry.
\end{abstract}

\maketitle

\vskip 0.5in

The space-time structure of the 
``fireball'' produced
in the collision between two heavy ions moving relativistically 
is studied by measuring  the two-particle momentum correlations
between pairs of identical particles. The quantum statistical 
effects of symmetrization cause an enhancement of the 
two-boson coincidence rate at small momentum differences
that can be related to the space-time size of the particle source.
This method, called HBT interferometry,  has been applied extensively 
in recent experiments at the Relativistic
 Heavy Ion Collider (RHIC) by the
STAR and PHENIX 
collaborations \cite{Kolb:2003dz}.

The invariant ratio of the 
 cross section for the production of two 
  pions of momenta $\bfp_1,\bfp_2$
to the product of single particle production cross sections is analyzed as
the correlation function $C(\bfp_1,\bfp_2)$.
 We define
 $\bfq$=$\bfp_1$--$\bfp_2$
and $\bfK$=$(\bfp_1$+$\bfp_2)/2$, with $\bfK_T$ as 
the component perpendicular to the beam direction. 
(We focus on mid-rapidity data, where 
$\bfK=\bfK_T$.)
The correlation function
is  parameterized
for small $\bfq$ as 
$
C(\bfq,\bfK)$ $\approx$$1 +
\lambda(1-R_o^2q_o^2-R_s^2q_s^2-R_l^2q_l^2),$
where $o,s,l$ 
represent directions parallel to $\bfK_T$, perpendicular
to $\bfK_T$ and the beam direction, 
and parallel to the beam direction\cite{BP-HBT}. 
Early \cite{Rischke:1996em} and recent \cite{Kolb:2003dz} hydrodynamic 
calculations predicted  that  a fireball evolving through a quark-gluon-hadronic phase 
transitions would emit pions
over a long time period, causing a large ratio $R_o/R_s$.
The puzzling experimental
result that $R_o/R_s\approx 1$ 
\cite{Adler:2001zd} is
part of what has been  
 called ``the RHIC HBT puzzle'' \cite{Heinz:2002un}. 

Data shows the medium produced by 200 GeV Au+Au collisions to be very dense.
Consequently, pions should emerge from
an opaque source\cite{Heiselberg:1997vh}. 
Analyticity tells us that opacity implies accompanying refractive effects.
Our purpose here is to derive and apply a relativistic quantum--mechanical
treatment of opacity and refractive effects that
simulataneously reproduces the values of $R_o, R_s, R_l$ and 
 the pion spectrum 
for central
RHIC Au+Au
collisions\cite{STARHBT,STARspec}.

The experimental observables depend on an emission function, 
the Wigner transform of the density matrix for the 
currents that emit pions. This emission function has often been
modelled (see  review \cite{Wiedemann:1999qn}),
 as a function  $S_0$ 
having  the form of a hydrodynamic source parameterization  
with approximately boost-invariant longitudinal 
dynamics: 
\bea
&&S_0(x,K)={\cal S}_0(\tau,\eta)\;B_\eta(\bfb, \bfK_T)/(2\pi)^{3}\label{fact}\\
&&{\cal S}_0(\tau,\eta)\equiv 
\frac{ \cosh\eta}{\sqrt{2\pi(\Delta \tau)^2}}  
            \exp \left[- {(\tau-\tau_0)^2 \over 2(\Delta \tau)^2}
           -{\eta^2\over2\Delta\eta^2}\right]\;\\
&&B_\eta(\bfb, \bfK_T)\equiv M_{T}{1\over \exp({K\cdot u-\mu_\pi \over T})-1  }
\rho(b),
\label{S0}\eea 
in which 
components of  $x^\mu$ are expressed using the
variables $\tau=\sqrt{t^2-z^2},~\eta=\frac{1}{2}\log{t+z\over t-z},$  $\bfb \equiv (x_1,x_2),$ and
  $K\cdot u= M_{T} \cosh\eta\cosh \eta_t(b) 
                - K_{T} \, \sinh \eta_t(b) \,\cos \phi$,
where $\phi$ is the angle between $\bfK_T$ and $\bfb$,
 $ M_T=\sqrt{K_{T}^2+m_\pi^2}$, and $\mu_\pi$ is the pion chemical potential.
The function $\rho(b)$ represents the cylindrically symmetric 
 transverse source density. We use
$\rho(b)= [1/(\exp((r-R_{WS})/a_{WS})+1)]^2$, reflecting superimposed nuclear densities.
      Eq.~(\ref{fact})
represents  the transverse flow rapidity using 
 a linear radial profile of strength $\eta_f$:
$
  \eta_t(b) = \eta_f {b \over R_{WS}}
$. Eq.~(\ref{fact}) incorporates the finite lifetime and size of the source:
 pions are
emitted for a duration controlled by the parameters $\Delta\tau$ and $\Delta\eta$.
Using Eq.~(\ref{fact}) is not sufficient  to capture all of the physics. As
the basis of the blast wave parameterization,  it gives $\Delta \tau\approx0$ and does not
predict the magnitude of the pion spectrum\cite{BW}. 

The salient feature 
of  the 200  GeV data is the high density of the produced matter, so we
treat  the effects of pion 
interactions with the dense medium.
We adopt  a single-channel  approach that 
uses 
the  
interaction--distorted incoming wave
$\Psi_{\bfp_1}^{(-)*}(x_1)$ 
in which\cite{GKW79}:
\bea  S(x,K) & = & \int d^4K' S_0(x,K')\int {d^4x'\over(2\pi)^4}\;e^{-i K'\cdot x'} \label{convo}  \\
&&\times ~\Psi_{\bfp_1}^{(-)}(x+x'/2) 
\Psi_{\bfp_2}^{(-)*}(x-x'/2).\nonumber\eea
One could imagine a generalization in which $\Psi$ would have two or more components 
(corresponding {\it e.g.} to a $\rho$ final state in addition to a pion final state). 
Our task here is 
to compute pion correlation functions, so that it is sufficient to use a single channel wave function.
One obtains the single-pion emission function from Eq.~(\ref{convo}) by 
using the same momentum (either $\bfp_1$ or
 $\bfp_2$) to compute the wave function $\Psi^{(-)}_\bfp$.

Using Eq.~(\ref{convo})  requires  evaluating  an     eight-dimensional integral,
and modelling the interactions that determine $\Psi_{\bfp_1}^{(-)}$. We use  
 symmetries to reduce the number of integrals and obtain a tractable treatment
of the interactions. First, note that 
$ \Psi_\bfp^{(-)}(x)$  
is an energy-eigenfunction \cite{GKW79}:
$ \Psi_\bfp^{(-)}(x)=e^{-i \omega_p\;x^0}\Psi_\bfp^{(-)}({\bf x}).$
 We assume that the 
matter formed in the central region of the collision 
is cylindrically symmetric with a very long axis, 
so that 
\bea
\Psi_{\bfp_{1,2}}^{(-)}({\bf x})& = &
e^{\mp iq_l z/2}\psi_{\bfp_{1,2}}^{(-)}({{\bf x_\perp}=\bfb}),\label{psidef}\\
\quad \bfp_{1,2} & = & \bfK\pm \bfq_T/2\pm {\bf\widehat{z}}\;q_l/2, \nonumber
\eea 
with $\psi_\bfp^{(-)}({\bfb})$  obtained by solving
a two-dimensional  Klein-Gordon  
equation
\bea \left(-\nabla_\perp^2 +U(b)\right)\psi_\bfp^{(-)*}({\bfb})=p^2
\psi_\bfp^{(-)*}({\bfb}).
\label{wave}\eea
The ``optical potential'' $U$ is a complex, azimuthally-symmetric function 
depending on pion momentum and local density that represents
the strength of the interaction between a pion and the medium.  Within our
formalism the influence of time-dependent effects in $U$ introduced by
the time-dependent source $S_0$ is incorporated in the energy dependence of
the optical potential. However, we note that the pions-medium interaction
time is 
restricted by $S_0$.

At large values of $K$ the solution of Eq.~(\ref{wave}) 
reduces to well-known semi-classical (eikonal) expression, but
our procedure is more general. For example
at $K$=0 there is no distinction between the out and side directions, so
$R_o$=$R_s$. This constraint is violated if the eikonal approximation to (\ref{wave}) is used. 
We use a partial wave expansion: 
 \bea
\psi_\bfp^{(-)*}({\bfb})=f_0(p,b)+2\sum_{m=1,\infty}f_m(p,b){(-i)}^m\;\cos{m\phi}\label{pwe}\eea
 to   solve Eq.(\ref{wave}) exactly  and maintain the constraint.

The optical potential accounts for 
situations in which the pion changes energy or disappears entirely due to its
interactions with the dense medium.
Suppose, {\it e.g.}, that 
the medium is a gas of pions. Then 
$\pi\pi$ scattering would be  the origin of $U$.
In the impulse approximation, the central optical potential would be
$U_0=-4\pi f\rho_0$, where $f$ is the complex 
forward scattering amplitude and $\rho_0$ the central density. 
For low energy pion-pion interactions, $4\pi$Im$[f(p)]=p\sigma,$
with $\sigma\approx 1\; $mb. At a 
momentum $p=1~\rm{fm}^{-1}$, 
using a pion density about
ten times the baryon density of ordinary nuclear matter, 
Im$[U(0)]\approx -0.15~\rm{fm}^{-2}$, representing significant opacity.
Furthermore, if two interacting 
pions each have less energy than half  of the rho meson mass, the 
final state interactions would be strongly attractive.

We  model the interaction $U$ for a situation in which 
the medium is dense, but we have not  otherwise specified  its  nature.
Now we consider chiral symmetry, which 
 gives a general form for the dispersion relation of low energy 
pions in nuclear matter:
 \bea
\omega^2=v^2(\hat{p}^2 +m^2_\pi(T)).\label{mode}\eea
 Here $m_\pi(T) $ is the $\pi$ screening mass, $v$ is the
so-called $\pi$ velocity, $\hat{p}$ is the momentum operator, and the
 pole mass is $m_\pi(T)v$. Son and Stephanov\cite{Son:2002ci}
argued   that $v$ and the pion pole mass 
decreases at 
temperatures near the critical temperature, while the screening
mass increases. Several papers have used specific low-order 
model calculations that challenge
 this result\cite{boy,sas}. Here we start with the general form (\ref{mode}) that appears
in both Refs.~\cite{Son:2002ci,boy} and use data to determine the parameters.  
We first define an equivalent $U$ by using
$\omega^2\equiv p^2+m_\pi^2$, Eq.~(\ref{mode}), and the Klein Gordon equation
$
\hat{p}^2 +U=p^2$,
to  obtain:
\bea U=-(m_\pi^2-v^2m_\pi^2(T)) -(1-v^2)\hat{p}^2. \label{uttt}\eea 

For matter of finite size, Eq.~(\ref{uttt}) suggests a
 potential with terms constant and proportional to $\hat{p}^2$. The latter 
has 
the form $w_2p^2 -\nabla\cdot w_K\nabla$, where $w_{2,K}$ are complex constants. 
For the 200 GeV data, 
 the results prefer a small value of $w_K$, so we take $w_K=0$.
The optical potential is therefore modelled as:
\bea
U_p(b)= -(w_0 + w_2 p^2)\rho(b) ,\label{uopt}
\eea 
with $w_0$ real (no opacity at p=0).
The density profile $\rho(b)$ is that specified in
$S_0(x,K)$. This simple form is sufficient to account
for the data we study. 

Eqns.(\ref{fact},\ref{uopt}) are the  essence of our model. 
These may be compared
with the Buda-Lund model, which is  an 
efficient representation of the data\cite{Buda} in which
 the temperature and fugacity are taken as position-dependent functions appearing
in a Boltzmann distribution. The effects of our optical potential could provide
an explanation of those deduced dependencies. 

The use of Eqs.~(\ref{psidef},\ref{fact}) in Eq.~(\ref{convo}) gives
\bea
S(x,K) & = & {1\over (2\pi)^2}{\cal S}_0(\tau,\eta) e^{iq^0t-iq_l z}
\int d^2b' \widetilde{B}_\eta(\bfb,\bfb')\nonumber\\
&& \times~\psi_{\bfp_1}^{(-)}(\bfb      +\bfb'     /2) 
\psi_{\bfp_2}^{(-)*}(\bfb      -\bfb'     /2),\label{cvo}\eea
where $
\widetilde{B}_\eta(\bfb,\bfb')\equiv \int d^2K'_{T}\;B_\eta(\bfb      ,\bfK'_{T}     )
\exp\left[-i\bfK_T'\cdot\bfb'     \right]. $
The range
of the variable $\bfb'$ appearing 
in $\widetilde{B}_\eta(\bfb,\bfb')$ 
is controlled by a size, $1/T$ that is much smaller than the source size. 
Thus it is reasonable to 
 ignore the $\pm\bfb'/2 $ appearing in (\ref{cvo}). 
Maintaining the phases is important,  
so we use 
$ \psi_\bfp^{(-)*}(\bfb\pm\bfb'/2)\equiv
e^{-i\bfp\cdot(\bfb       \pm\bfb'/2)}
\chi_\bfp(\bfb\pm\bfb'/2)\to 
e^{-i\bfp\cdot(\bfb\pm\bfb'/2)}\chi_\bfp(\bfb)$.
This approximation is exact in several different limits: plane wave, 
short wavelength, 
 long wave length, 
and taking   $(R_{WS}T\to\infty)$.

We expand the exponentials in (\ref{cvo}) involving $q^0,q_l$ to 2nd order,
 expand the Bose-Einstein function
in a series of Boltzmann functions of temperature $T_n=T/n$, and analytically
evaluate integrals over   
 $\tau$ and $\eta$ 
  to obtain:
\bea  &&C(\bfq,\bfK_T)  =  1-q_l^2R_l^2 -q^2_o\beta^2\widetilde{\Delta\tau}^2+
\frac{\vert\Phi_{12}\vert^2}{\Phi_{11}
\Phi_{22}}, \\
&&\Phi_{ij}  =\sum_n  \int d^2bf_0(\xi_n(b)) 
\psi^{(-)}_{\bfp_i}(\bfb)\psi_{\bfp_j}^{(-)*}(\bfb)
B_n(\bfb,\bfK_T)\nonumber \\
&&B_n(\bfb,\bfK_T)=\exp({\mu_\pi+K_T\sinh\eta_t(b)\cos\phi\over T_n})M_T\rho(b)\nonumber\\
&&R_l^2  =  {(3\Delta\tau^2+\tau_0^2)F_1(K)\over F_0(K)}\nonumber\\
&&\widetilde{\Delta\tau}^2=(3\Delta\tau^2+\tau_0^2){F_3(K_T)\over F_0(K_T)}-
\left\vert
{(\tau_0^2+\Delta\tau^2) F_2(K_T)\over \tau_0F_0(K_T)}\right\vert^2\nonumber\\
&&F_m(K)  =\sum_n  \int d^2bB_n(\bfb,\bfK_T)
f_m(\xi_n(b)){{\vert}}\psi^{(-)}_{\bfK_T}(\bfb){{\vert}}^2
\nonumber
\eea
where $\xi_n(b)=M_T\cosh\eta_t(b)/T_n+1/\Delta\eta^2,\;
f_1(\xi)\equiv 2 K_0(\xi)/\xi+4K_1(\xi)/\xi^2,\;
f_0(\xi)\equiv 2K_1(\xi),\;f_2(\xi)=K_0(\xi)+K_2(\xi),\;f_3(\xi)=
2(K_1(\xi)+K_2(\xi)/\xi)$, 
and $K_j$ are modified Bessel functions.
The pion spectrum is  given by:
\bea
\left\langle
{dN\over 2\pi M_{T}  dM_{T} dY}\right\rangle_{\vert Y\vert<0.5}
=~{\tau_0\over 8\pi^3}e^{1\over\Delta\eta} F_0(\bfK_T).\label{spectrum}\eea
The angular  integrals are  performed 
analytically  
Eq.~(\ref{pwe}). 
 The transverse HBT radii are 
 $R_i(K_T)=\sqrt{1-C(\Delta q_i,K_T)}/\Delta q_i$, with $i=o,s$ and $\Delta q_i\approx K_T/40$.

The values of the parameters 
$T,\eta_f,\Delta\tau,R_{WS},a_{WS},$
$w_{0,2},\tau_0,
\Delta \eta$ and $\mu_\pi$ 
are varied to reproduce the STAR 200 GeV data for $R_o, R_s, R_l$ \cite{STARHBT} and
the magnitude and shape of the pion spectrum \cite{STARspec}.
The resulting parameters are
displayed in Table I, with the variances that produce a $\chi^2$ increase of one unit.  
The agreement of the curves with STAR HBT radii \cite{STARHBT} in Fig. 1 
and with the STAR pion spectrum \cite{STARspec} in Fig. 2
 is quite good, giving a $\chi^2\approx$ 3.7 per data point and 5.6 per degree of freedom.

\begin{figure}
\includegraphics[width=10 cm]{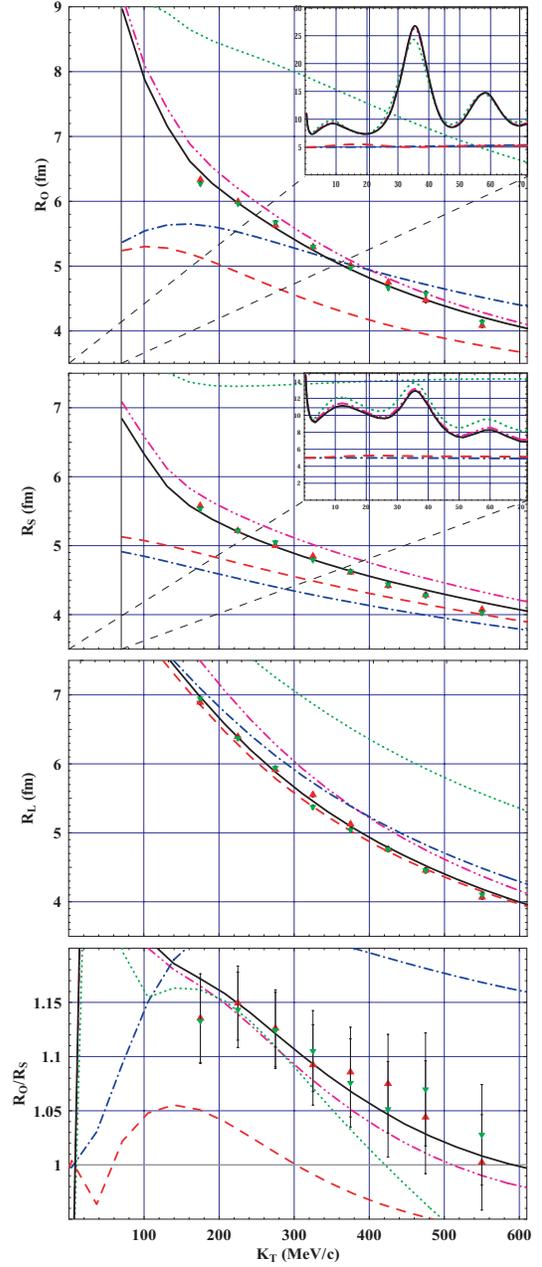}
\caption{\label{one}  (Color online) HBT Radii $R_s, R_o,$, $R_l$ and the ratio $R_o/R_s$; 
Data \cite{STARHBT}): $\nabla$ (green) $\Rightarrow \pi^{+}\pi^{+}$; $\triangle$ (red) $\Rightarrow \pi^{-}\pi^{-}$.  
Curves: solid (black) $\Rightarrow$ full calculation; 
dotted (green) $\Rightarrow\eta_f=0$ (no flow); dashed (red) $\Rightarrow$ Re$[U]$=0 (no refraction); dot-dashed (blue) $\Rightarrow$ 
$U$=0 (no potential), double-dot-dashed (magenta) $\Rightarrow$ substituting Boltzmann for Bose-Einstein thermal distribution.
Insets show predictions of low-$K_T$ resonance behavior in $R_o$ and $R_s$.}
\end{figure}

\small
\begin{table*}
  \centering
  \caption{Parameters of the calculation with variances}
  \vspace{0.1cm}
  \begin{tabular}{|rrrrrllrrr|}\hline
   $~~T(MeV)$ & $~~\eta_f~~$ & $\Delta\tau(fm/c)$ & $R_{WS}(fm)$ & $~~a_{WS}(fm)$& $~~w_0(fm^{-2})$& $~~~~~~~~w_2$
& ~~~$\tau_{0}(fm/c)$ &  ~~~$\Delta\eta$~~~ & $\mu_\pi(MeV)$\\
\hline
173.2~~   & ~~1.314 & 2.852~~~ & 11.728~ & 0.725~~ & ~~~0.137 & 0.582~+~$i$~0.121 & 8.23~~~ &  1.063~~~& 123.2~~~\\
~~$\pm$1.6~~   & ~~$\pm$0.025 & $~\pm$0.067~~~ & ~$\pm$0.056~~~ & $\pm$0.015~~ & ~$\pm$0.046 & $\pm$0.014~~$\pm$0.002~~~ & ~$\pm$0.10~~~ &  $\pm$0.032~~~& $\pm$1.1~~~\\

      \hline
      \end{tabular}
      \end{table*}
\normalsize

\begin{figure}
\includegraphics[width=8.5 cm]{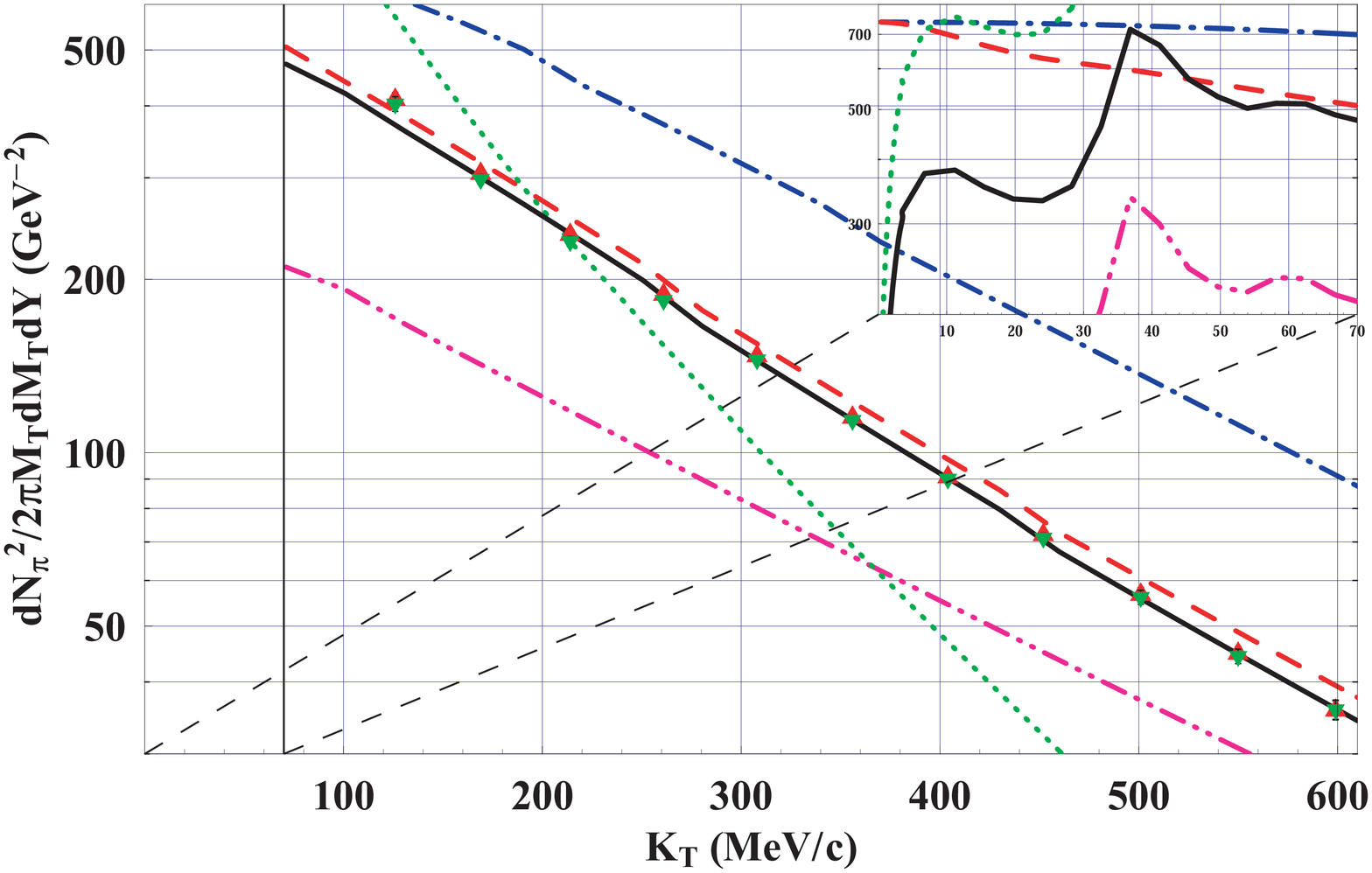}
\caption{\label{three} (Color online) 
Pion momentum spectrum.
Data \cite{STARspec}: $\nabla$ (green) $\Rightarrow \pi^{+}$; $\triangle$ (red) $\Rightarrow \pi^{-}$. 
Inset is low-$K_T$ prediction.}

\end{figure}

We need to assess and interpret the values of our parameters.
The temperature ($T$=173 MeV) is comparable to the $T_c$ expected for a chiral phase
 transition ($\approx$160 MeV).
The  
flow rapidity ($\eta_f$=1.31)  corresponds to a maximum flow velocity of 0.85 c,
 which is not unreasonable.  The source size is large ($R_{WS}$=11.7 fm),
 4.4 fm larger than a gold nucleus ($R$=7.3 fm).  If the system
 requires an expansion time $\tau_0$= 8.22 fm/c to expand by this
 amount, the average expansion velocity would be about 0.5 c, which is reasonable. 
 The emission duration ($\Delta\tau$=2.9 fm/c) is comfortably smaller than $\tau_0$. The longitudinal width ($\Delta\eta$=1.063) is smaller than expected, but
 is large enough that
the system's axial length ($\sim 2\tau_0 \Delta \eta\sim 17.5$ fm) is sufficient for the
approximate validity of Eq.~(\ref{psidef}). 

We assess the strength of $U$ by examining it at $p \approx 1$ fm$^{-1}$. At the system center, the
 momentum-dependent term of $U$
is --(0.58+0.12$i$) fm$^{-2}$. To understand the imaginary part, which is comparable to
 our estimate given above (and consistent with the presence of high density
matter), consider the equivalent classical mean free path: 
 $p/\vert Im[U]\vert \approx 8~\rm{fm}\ll 2R_{WS}$. 
Thus, the imaginary potential is large enough to 
restrict emission of pions from regions 
deep inside the medium. To understand the real part,
note that the strength of the attraction is greater than $m_\pi^2=0.49 \;{\rm fm}^{-2}$. 
Thus inside the medium,  the pion acts as if it has no mass. This is why we
assert that a chiral phase transition has occured. 

We further 
recall Eq.~(\ref{uttt}), and 
note that a momentum-dependent term 
 of -0.58 fm$^{-2}$ corresponds $v= 0.65$. The
 momentum independent term of $U$ (-.14 fm$^{-2}$) corresponds to 
of $vm_\pi(T)=0.6$ fm$^{-1}$. These values are comparable to the estimates of \cite{Son:2002ci},
but not too far from the lowest value $v=0.83$ of Ref.~\cite{sas}.

The chemical potential and Bose-Einstein distribution have modest effects on the radii
 but are very important for the normalization and slope of the pion spectrum.
 

Figs.~\ref{one},~\ref{three} show the importance of both the real and imaginary parts of the
optical potential: omitting either drastically changes the predictions. 
The attractive  real potential is the critical element needed to reproduce the
$K_T$ dependence of $R_o$ and $R_S$.

We predict peaks in $R_o$ and $R_s$ at low momentum ($p\approx$ 15--65 MeV/c)
and  a rapid rise and peaking in the low-momentum spectrum.  
This is a pionic version of the
Ramsauer effect, in which cross sections show peaks when the scattered wave
is in phase with the incident plane wave. We 
confirm the computed existence of such peaks through 
analytic calculations for a purely real
attractive square well potential. 
The PHOBOS detector 
(and perhaps the BRAHMS detector) at RHIC could,
in principle,
confront these low--momentum predictions of structure.

Our 
relativistic quantum-mechanical treatment of refractive and opacity effects
on two-pion correlations and $\pi$ spectra
has surmounted a number of technical problems 
associated with previous semi-classical approximations, while achieving an 
excellent description of the 200 GeV STAR data. The results are consistent with the
interpretation that the dense medium has undergone a chiral phase transition.  
The
definitive predictions of interesting momentum dependences at small
 momenta should be testable in present and future RHIC
experiments.

This work is partially supported by the USDOE grants DE-FG-02-97ER41014 and DE-FG-02-97ER41020. GAM thanks LBL, TJNAF and BNL for their
hospitality during the course of this work. We thank J.~Draper, S.~Reddy and D. Son for useful
discussions. 

\end{document}